\documentclass[10pt]{article}

\usepackage{setspace}
\usepackage[dviwin]{graphicx}
\usepackage{amsfonts}
\usepackage{amssymb}
\usepackage{mathrsfs}
\usepackage{amsmath}
\usepackage{epsfig}
\usepackage{relsize}
\usepackage{graphicx}
\usepackage{mathabx}
\usepackage{caption}

\usepackage{dsfont}
\usepackage{mathrsfs}
\usepackage{esint}
\usepackage{textpos}
\usepackage{wrapfig}
\usepackage{lipsum}
\usepackage{hyperref}
\hypersetup{
    colorlinks = true,
    linkcolor = [rgb]{0 .4 .8},
    citecolor=[rgb]{1 .2 .2}
}
\usepackage{array}

\newcommand{\bea}{\begin{equation}}
\newcommand{\eea}{\end{equation}}
\newcommand{\bear}{\begin{eqnarray}}
\newcommand{\eear}{\end{eqnarray}}
\newcommand{\bearr}{\begin{eqnarray*}}
\newcommand{\eearr}{\end{eqnarray*}}

\newcommand{\beal}{\begin{align}}
\newcommand{\eeal}{\end{align}}
\newcommand{\beall}{\begin{align*}}
\newcommand{\eeall}{\end{align*}}

\newcommand{\CP}{\mathds{C}\mathds{P}}
\newcommand{\CC}{\mathds{C}}

\newcommand{\dd}{\partial}

\newcommand{\comment}[1]{}

\usepackage{xcolor}
\usepackage{empheq}

\makeatletter
\def\@seccntformat#1{\@ifundefined{#1@cntformat}%
{\csname the#1\endcsname\quad}
{\csname #1@cntformat\endcsname}
}
\def\section@cntformat{{\normalfont\large\thesection.}\quad}
\def\subsection@cntformat{\textsection\, \thesubsection.\quad}
\def\subsubsection@cntformat{\textsection\textsection\, \thesubsubsection.\quad}
\makeatother

\newcommand{\ssection}[1]{%
  \section[#1]{\centering\normalfont\scshape #1}}
\newcommand{\ssubsection}[1]{%
   \subsection[#1]{\raggedright\normalfont  #1}}
\newcommand{\ssubsubsection}[1]{%
   \subsubsection[#1]{\raggedright\normalfont  #1}}

\newsavebox\MBox

\begin{document}

\title{Classical solutions of a flag manifold $\sigma$-model}
\author{Dmitri Bykov\footnote{Emails:
dmitri.bykov@aei.mpg.de, dbykov@mi.ras.ru}  \\ \\
{\small $\bullet$ Max-Planck-Institut f\"ur Gravitationsphysik, Albert-Einstein-Institut,} \\ {\small Am M\"uhlenberg 1, D-14476 Potsdam-Golm, Germany} \\ {\small $\bullet$ Steklov
Mathematical Institute of Russ. Acad. Sci.,}\\ {\small Gubkina str. 8, 119991 Moscow, Russia \;}}
\date{}
\maketitle
\vspace{-0.5cm}
\begin{center}
\line(1,0){370}
\end{center}
\vspace{-0.2cm}
\textbf{Abstract.} We study a $\sigma$-model with target space the flag manifold $U(3)\over U(1)^3$. A peculiarity of the model is that the complex structure on the target space enters explicitly in the action. We describe the classical solutions of the model for the case when the worldsheet is a sphere $\CP^1$.
\vspace{-0.7cm}
\begin{center}
\line(1,0){370}
\end{center}

\vspace{0.3cm}
In this paper we will solve the equations of motion (e.o.m.) of the $\sigma$-model proposed in \cite{Bykov} (reviewed in Sec. \ref{flag}) for the case when the worldsheet $\mathscr{M}$ is a sphere $\CP^1$. The target space of the model is the manifold of full flags in $\CC^3$, which we will denote by $\mathcal{F}_3$. It can be viewed as the space of ordered triples of orthogonal lines in $\CC^3$ passing through the origin, and is also representable as a quotient space:
\bea\label{quot}
\mathcal{F}_3=\frac{U(3)}{U(1)^3}\,.
\eea
From the structure of the quotient (\ref{quot}) it is clear that there are three natural forgetful maps: $\{ \pi_i: \mathcal{F}_3\to \CP^2, \;i=1, 2, 3 \}$. For this reason the properties of the flag manifold are tightly related to the properties of the underlying $\CP^2$'s. As we shall see, solutions to the flag $\sigma$-model e.o.m. are to a large extent expressible through the solutions of the $\CP^2$ model. Due to this, and to introduce the notations, we begin by defining the $\sigma$-model with target space $\CP^2$.

\section{The $\CP^2$ $\sigma$-model}

We will be thinking of $\CP^2$ as the quotient $\CP^2=(\CC^3-\{0\})/\CC^\ast.$
A map $v:~\mathscr{M}~\to~\CP^2$ from a Riemann surface $\mathscr{M}$ can be described by a vector-valued function $v(z, \bar{z}) \in~\CC^3$, where $z, \bar{z}$  are coordinates on the worldsheet $\mathscr{M}$. We may assume that the vector $v$ is in fact normalized, that is $v\in S^5\subset \CC^3$: $\sum\limits_{i=1}^3\,|v_i|^2:=\bar{v}\circ v=1$, and henceforth we will use this normalization. This is a partial gauge for the gauge group $\CC^\ast$, which breaks it down to $U(1)$.

Introduce the covariant derivative
\bea\label{covdiv}
D_i^{(v)} w:=\dd_i w-q_w\cdot (\bar{v}\circ \dd_i v)\, w\,,\quad\quad i=\{z, \bar{z}\}
\eea
where $q_w$ is the $U(1)$-charge of $w$, normalized so that $q_v=1$.
In most of the applications of (\ref{covdiv}) below $w$ is a vector obtained by applying covariant derivatives to the basic map $v$ or its conjugate $\bar{v}$. For example, $w=\{v, D^{(v)}_z v, D^{(v)}_{\bar{z}}v, D^{(v)}_z D^{(v)}_{\bar{z}}v, \ldots \}$, in which case $q_w=1$, or $w=\{\bar{v}, D^{(v)}_z \bar{v}, D^{(v)}_{\bar{z}}\bar{v}, D^{(v)}_z D^{(v)}_{\bar{z}}\bar{v}, \ldots \}$, in which case $q_w=-1$. When this does not lead to confusion, we will sometimes simply write $D_z$ in place of $D_z^{(v)}$, $D_{\bar{z}}$ for $D_{\bar{z}}^{(v)}$.

The covariant derivative has the Leibniz property:
$
D_i^{(v)}(a\cdot b)=D_i^{(v)}(a)\cdot b+a\cdot D_i^{(v)}(b)\, .
$
The commutator of covariant derivatives produces the pull-back of the Fubini-Study form:
\bea\label{comm}
[D_z^{(v)}, D_{\bar{z}}^{(v)}]=D_{\bar{z}}^{(v)}\bar{v}\circ D_z^{(v)} v-D_z^{(v)}\bar{v}\circ D_{\bar{z}}^{(v)}v\;.
\eea

The action of the $\CP^2$ $\sigma$-model (with zero $\theta$-term) is:
\bea
\mathcal{S}=\int\limits_{\mathscr{M}}\,{i\over 2}\,dz\wedge d\bar{z}\,\big(\|D_z v\|^2+\|D_{\bar{z}}v\|^2\big)
\eea
The equation of motion following from this action reads
\bea\label{CP2eom}
D_{\bar{z}}^{(v)}D_z^{(v)} v =\alpha \,v,
\eea
where $\alpha$ is a scalar function. Multiplying this equation by $\bar{v}$ and using the Leibniz property of the covariant derivative together with the identity $\bar{v}\circ D_z^{(v)}v=0$ (which follows from the definition (\ref{covdiv})), we find that
$
\alpha=-\,\|D_z v\|^2
$.
Since, according to (\ref{comm}), $[D_z, D_{\bar{z}}]$ is a scalar function, the equation (\ref{CP2eom}) can be equivalently rewritten as
\bea\label{harm2}
D_z^{(v)} D_{\bar{z}}^{(v)} v =\tilde{\alpha} \,v
\eea
A map $v$ satisfying (\ref{CP2eom})-(\ref{harm2}) is called \emph{harmonic}. For a review of the theory of harmonic maps we refer the reader to \cite{Salamon}.

\section{The flag manifold $\sigma$-model}\label{flag}

As mentioned earlier, we wish to consider in detail the $\sigma$-model introduced in \cite{Bykov}, which we will recall momentarily. In that case the target space is the flag manifold $\mathcal{F}_3~=~\frac{U(3)}{U(1)^3}$, parametrized by the orthonormalized vectors $u_i$ ($u_i \circ \bar{u}_j=\delta_{ij}$), modulo phase rotations $u_k \to e^{i \alpha_k}\,u_k$. Introduce the currents
\bea
J_{mn}:=u_m\circ d\bar{u}_n\;,\quad\quad m, n=1, 2, 3.
\eea
The off-diagonal currents $\{J_{mn},\; m\neq n\}$ comprise the vielbein (and are defined up to phase factors).
Note that $J_{nm}=-\bar{J}_{mn}$. One can define an \emph{almost complex structure} on $\mathcal{F}_3$ by picking any three mutually non-conjugate forms, $J_{m_1 n_1}, J_{m_2 n_2}, J_{m_3 n_3}$, and declaring them holomorphic. The other three, being conjugate to these, are therefore antiholomorphic. In order to decide, which of these complex structures are integrable, a diagrammatic representation is useful. Draw three nodes and directed arrows from node $m_1$ to $n_1$, $m_2$ to $n_2$ and $m_3$ to $n_3$. Integrability of the so-defined complex structure is equivalent to the condition that the graph is acyclic (i.e. does not have a directed closed loop). Let us prove this. First of all, let $e_m, \;m=1, 2, 3$ be the standard unit vectors with components $(e_m)_n=\delta_{mn}$. To the holomorphic one-forms one can associate a subspace $m_+$ of the Lie algebra $(\mathfrak{su}(3))_\CC=\mathfrak{sl}(3)$ as follows:
\bea
m_+=\mathrm{Span}(E_{m_1 n_1}, E_{m_2n_2}, E_{m_3n_3}),\quad \textrm{where}\quad E_{mn}=e_m\otimes e_n
\eea
Integrability of the complex structure is equivalent to the requirement that $m_+$ is a subalgebra: $[m_+, m_+]\subset m_+$. On the other hand, the matrices $E_{mn}$ have the commutation relations
\bea
[E_{mn}, E_{pq}]=\delta_{np}E_{mq}-\delta_{mq} E_{pn}
\eea
Remembering that $E_{mn}$ is represented by an arrow from $m$ to $n$, one sees that the closedness of $m_+$ under commutation is equivalent to the following statement:
\bear\label{property}
&&\textrm{For any two consecutive arrows $m \to n$ and $n \to p$}\\ \nonumber
&&\textrm{their `shortcut' segment $(m, p)$ has the arrow $m\to p$}
\eear
For the diagram with three vertices, i.e. for the $\mathfrak{su}(3)$ case under consideration, it is clear that the cyclic quivers are the only ones that are ruled out.

\begin{figure}
    \centering 
    \includegraphics[width=0.4\textwidth]{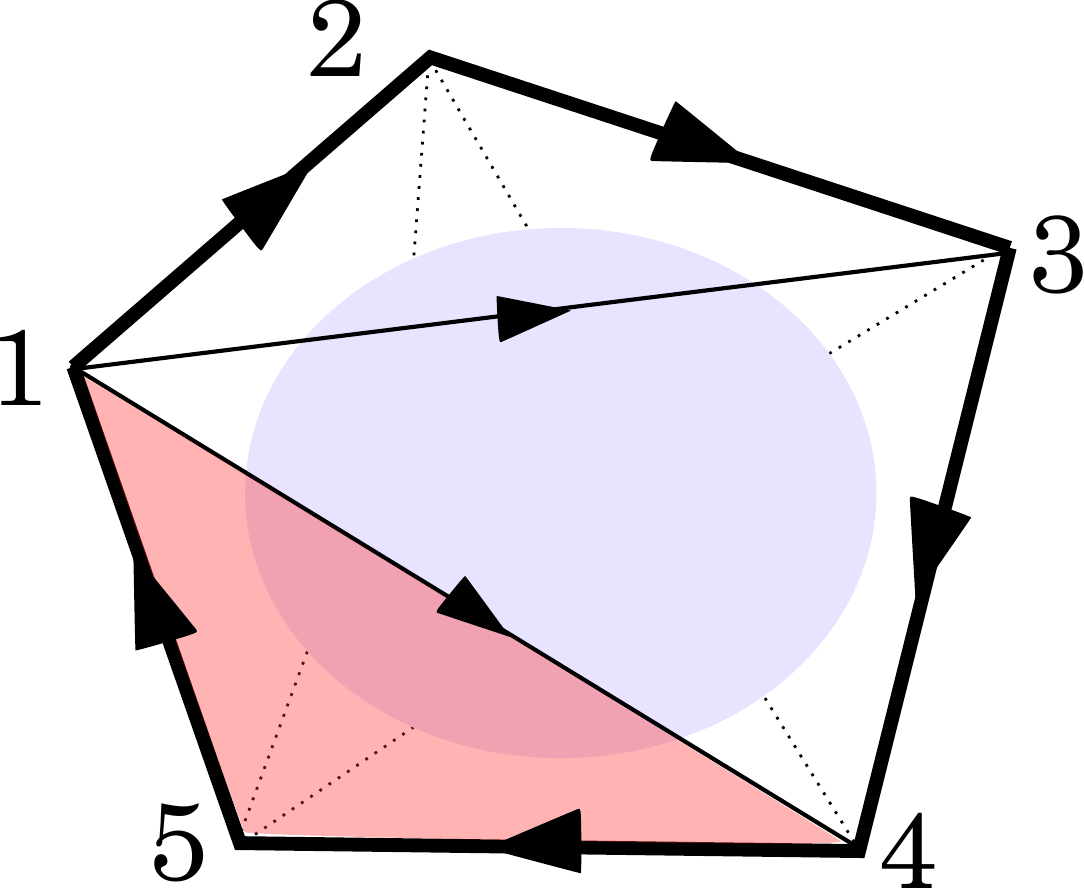}
  \caption{The procedure showing that a cycle $(1,2,3,4,5)$ in a graph leads to the violation of condition (\ref{property}). Using (\ref{property}), we replace the pair of segments $(1,2), (2,3)$ by $(1,3)$, i.e. cut a corner. Then we replace $(1,3)$, $(3,4)$ by $(1,4)$, arriving at the cyclic red triangle, which violates (\ref{property}).}
  \label{cut}
\end{figure}

In the general case, corresponding to the flag manifold $U(N)\over U(1)^N$, suppose we have $N$ pairwise-connected vertices, and the graph is acyclic. Then the requirement (\ref{property}) is satisfied, since otherwise there would be a cycle with three vertices. Reversely, suppose the graph has a cycle. Then, using (\ref{property}), one can `cut corners' to reduce again to the cycle with three vertices, which is prohibited (see Fig. \ref{cut}).

\begin{figure}[h]
    \centering 
    \includegraphics[width=0.65\textwidth]{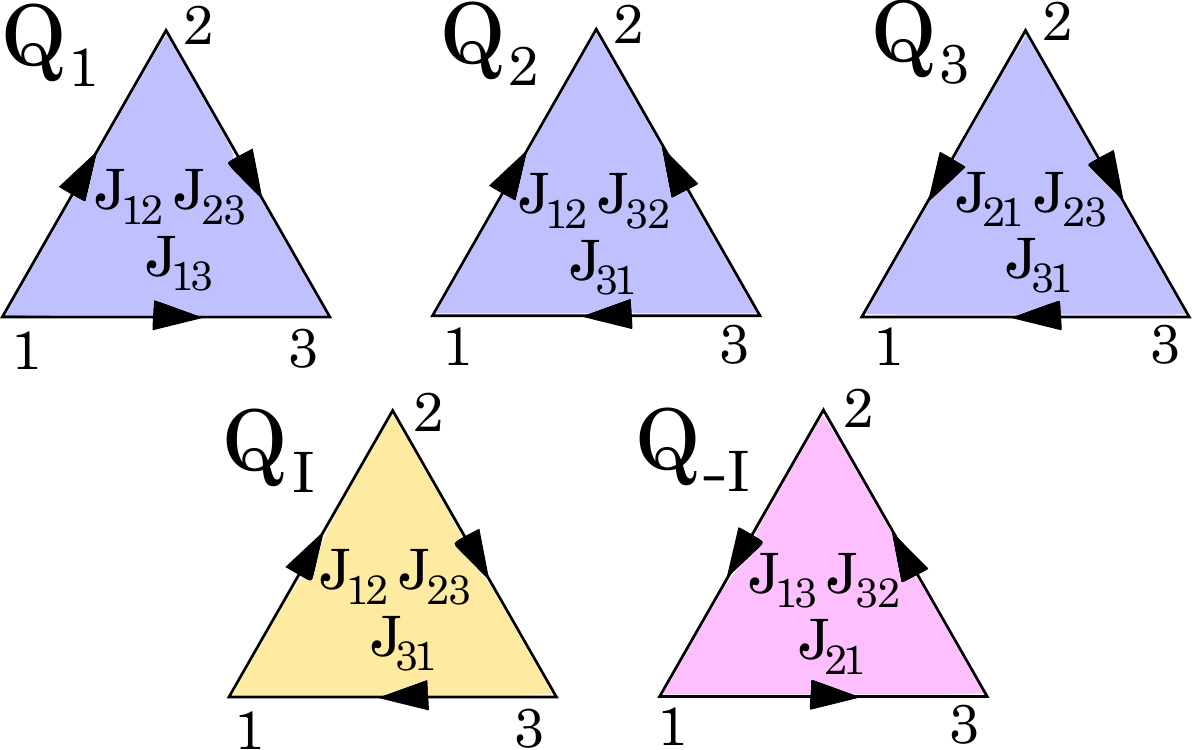}
    \caption{The triangles indicate the complex structures, whose associated holomorphic curves are solutions of the $\sigma$-model. The three top triangles correspond to integrable complex structures, whereas the two lower ones correspond to the non-integrable ones.}
\label{cstr}
\end{figure}

We now return to the $\mathfrak{su}(3)$ case. Once we are given an acyclic quiver $\mathcal{Q}$, the action proposed in \cite{Bykov} is 
\bea\label{action}
\mathcal{S}_{\mathcal{Q}}=\int\limits_{\mathscr{M}}\,{i\over 2}\,dz\wedge d\bar{z}\,\left(\sum\limits_{\begin{subarray}{l}\textrm{Arrows}\\
    \;m\rightarrow n\end{subarray}}\; \big|(J_{mn})_{\bar{z}}\big|^2\right)
\eea
It was also shown that the actions corresponding to three different integrable complex structures, whose associated quivers are shown in Fig. \ref{cstr} in blue, differ only by topological terms:
\bea\label{diff}
\mathcal{S}_{\mathcal{Q}_1}-\mathcal{S}_{\mathcal{Q}_2}=\mathrm{const.}\quad\quad \textrm{etc.},
\eea
Therefore they produce the same e.o.m. In particular, it follows from (\ref{action})-(\ref{diff}) that a curve, holomorphic in a complex structure corresponding to one of the three quivers $\mathcal{Q}_1, \mathcal{Q}_2, \mathcal{Q}_3$ in Fig. \ref{cstr}, is a solution to the e.o.m. What is more surprising, however, is that a curve holomorphic in any of the two non-integrable complex structures is a solution to the e.o.m. as well. To see this, one needs to write out the e.o.m. explicitly:
\bea\label{eomflag}
\mathscr{D}_z (J_{12})_{\bar{z}}=0,\quad\quad \mathscr{D}_{z}(J_{31})_{\bar{z}}=0,\quad\quad \mathscr{D}_z (J_{23})_{\bar{z}}=0\quad\quad \textrm{and c.c. ones}
\eea
Here $\mathscr{D}$ is the $U(1)^3$-covariant derivative, acting as follows: $\mathscr{D}J_{mn}:=d J_{mn}+(J_{mm}-J_{nn})\wedge J_{mn}$. One sees that $(J_{12})_{\bar{z}}=(J_{31})_{\bar{z}}=(J_{23})_{\bar{z}}=0$ is a solution to (\ref{eomflag}), and this is precisely the defining equation of a curve, holomorphic in the almost complex structure that corresponds to the cyclic quiver $\mathcal{Q}_I$ in Fig. \ref{cstr}. As regards the opposite non-integrable complex structure $-I$, one can rewrite the equations (\ref{eomflag}) alternatively as
\bea\label{eomflag2}
\mathscr{D}_{\bar{z}} (J_{12})_{z} \sim (J_{13}\wedge J_{32})_{z\bar{z}},\quad \mathscr{D}_{\bar{z}}(J_{31})_{z}\sim (J_{32}\wedge J_{21})_{z\bar{z}},\quad  \mathscr{D}_{\bar{z}} (J_{23})_{z}\sim (J_{21}\wedge J_{13})_{z\bar{z}}
\eea
In the complex structure $-I$ the l.h.s. vanishes, and all of the one-forms in the r.h.s. are of type $(1, 0)$ (i.e. proportional to $dz$), hence their wedge products vanish as well. Note, however, that the e.o.m. written with reference to the complex structures $I$ (\ref{eomflag}) and $-I$ (\ref{eomflag2}) are of rather different form (despite being equivalent), which is the reason that we present the corresponding quivers in Fig. (\ref{cstr}) in different color.

\vspace{0.3cm}
\textbf{Remark.} In order to understand the integrable complex structures on $\mathcal{F}_3$, it is most useful to recall the following definition of the flag manifold (see, for example,~\cite{Lawson}):
\bea\label{flagcompl}
\mathcal{F}_3=\{w_0v_0+w_1v_1+w_2v_2=0,\;\;(w, v) \in \CP^2\times \CP^2\}
\eea
Such an embedding into $\CP^2 \times \CP^2$ defines a complex structure on $\mathcal{F}_3$. In order to make contact with our previous definitions in terms of the one-forms $J_{mn}$, consider for instance the complex structure corresponding to the quiver $\mathcal{Q}_1$, and a curve $\mathscr{C}$ holomorphic in this complex structure. The following facts are easily derived:
\bear
\bar{u}_2 \circ \dd_{\bar{z}}u_1=0,\quad \bar{u}_3 \circ \dd_{\bar{z}} u_1=0\quad \Rightarrow\quad D_{\bar{z}}^{(u_1)} u_1=0\\
u_1 \circ  \dd_{\bar{z}} \bar{u}_3=0,\quad u_2 \circ  \dd_{\bar{z}} \bar{u}_3=0\quad \Rightarrow\quad D_{z}^{(u_3)} u_3=0
\eear
This means that the projections of $\mathscr{C}$ to the $\CP^2$'s with coordinates $u_1, \bar{u}_3$ are holomorphic curves. Moreover, $u_1 \circ \bar{u}_3=0$. Comparing with (\ref{flagcompl}), one realizes that $(w, v)$ in (\ref{flagcompl}) may be identified with $(u_1, \bar{u}_3)$. All other integrable complex structures on $\mathcal{F}_3$ are obtained by replacing $(w, v)$ with the various pairs $(u_i, \bar{u}_j)$ and using the embedding (\ref{flagcompl}).

\section{Critical maps $\CP^1\to \mathcal{F}_3$}

We call a map $\mathscr{M}\to \mathcal{F}_3$ \emph{critical} if it is a solution of the e.o.m. (\ref{eomflag}). Henceforth in this paper we will be concerned with the case $\mathscr{M}=\CP^1$.

From the equations (\ref{eomflag}) one deduces the following conservation equation:
\bea\label{consflag}
\dd_{z}\left((J_{12})_{\bar{z}}(J_{23})_{\bar{z}}(J_{31})_{\bar{z}}\right)=0
\eea
Note that the expression in brackets is a section of the cube of the canonical bundle $K$ of $\CP^1$, and the conservation law states that it has to be anti-holomorphic, i.e. $(J_{12})_{\bar{z}}(J_{23})_{\bar{z}}(J_{31})_{\bar{z}} \in H^0(K^3, \CP^1)$. However, as $H^0(K^3, \CP^1)=~0$, the only such section is zero. Hence
\bea\label{noholdif}
(J_{12})_{\bar{z}}(J_{23})_{\bar{z}}(J_{31})_{\bar{z}}=0
\eea
Suppose
\bea\label{J31}
(J_{31})_{\bar{z}}=0,
\eea
then the remaining equations (\ref{eomflag}) assume the form
\bea\label{u2harm}
\bar{u}_1\circ D_{\bar{z}}D_z u_2=0,\quad\quad  \bar{u}_3\circ D_{\bar{z}}D_z u_2\quad \Rightarrow\quad D_{\bar{z}}D_z u_2=\alpha\,u_2,
\eea
where $\alpha$ is an arbitrary (scalar) function. Hence $u_2(z, \bar{z})$ is harmonic (see (\ref{CP2eom})).

\ssubsection{Harmonic maps $\CP^1\to \CP^2$}

In this section we review the construction of the harmonic maps $\CP^1 \to \CP^2$, which was carried out long ago \cite{Din} (using a method developed in \cite{Calabi} for the description of minimal maps $S^2\to S^n$). The key property of such maps, which lies at the heart of the construction, is called `complex isotropy':
\bea\label{isotropy}
D_z^n \bar{v} \circ D_z^m v=0\quad\quad\textrm{for}\quad m+n>0
\eea
Note that this property does not hold, in general, for harmonic maps $\mathscr{C}_g \to \CP^2$, where $\mathscr{C}_g$ is a curve of positive genus $g>0$.

\vspace{0.3cm}
\textbf{Proof of (\ref{isotropy})}.

First of all, by definition of covariant derivative,
\bea\label{ortho1}
\bar{v}\circ D_z v=0= D_z \bar{v}\circ v
\eea
Suppose we have proven (\ref{isotropy}) for $m+n \leq N$. Then it follows that $D_z^{n+1} \bar{v} \circ D_z^m v$ for $m+n=N$ is a tensor under complex-analytic changes of variables. Indeed, under a coordinate change $z=z(w)$ one has
\bea
D_z^{n+1} \bar{v} \circ D_z^m v \to \left({\dd w \over \dd z}\right)^{n+m+1}\,D_w^{n+1} \bar{v} \circ D_w^m v+\sum\limits_{m+n < N}\,g_{m,n}\, D_w^{n+1} \bar{v} \circ D_w^m v,
\eea
and the sum vanishes by our assumption. Therefore $D_z^{n+1} \bar{v} \circ D_z^m v \in \Gamma(K^{m+n+1}, \CP^1)$. Consider
\bea
\dd_{\bar{z}}(D_z^{n+1} \bar{v} \circ D_z^m v)=(D_{\bar{z}}D_z^{n+1} \bar{v} )\circ D_z^m v+D_z^{n+1} \bar{v} \circ (D_{\bar{z}} D_z^m v)
\eea
for $m+n=N$. Using the commutation relation (\ref{comm}) for covariant derivatives and the harmonicity of $v$, we find that $D_{\bar{z}} D_z^m v=\sum\limits_{k<m}\,f_k D_z^k v$, and we have already proven that $D_z^{n+1} \bar{v}\circ D_z^k v=0$ for $n+k<N$. 

Therefore $\dd_{\bar{z}}(D_z^{n+1} \bar{v} \circ D_z^m v)=0$, so that $D_z^{N-m+1} \bar{v} \circ D_z^m v$ is a holomorphic section of the line bundle $K^{N+1}$ over $\CP^1$. The key property  (which we already used above for the case $m=3$) is that
\bea
H^0(K^{m}, \CP^1)=0\quad\quad \textrm{for}\quad m>0,
\eea
hence such a section is necessarily zero, leading to (\ref{isotropy}).
$\blacksquare$

\vspace{0.3cm}
Once (\ref{isotropy}) is established, consider the following sequence of maps:
\bea\label{backlund}
\ldots \rightarrow D_{\bar{z}}^2 v\rightarrow D_{\bar{z}}v\rightarrow v \rightarrow D_z v \rightarrow D_z^2 v \rightarrow \ldots
\eea
The sequence can be continued to the left and right, however for $\CP^2$ it is sufficient to consider the terms shown in (\ref{backlund}). Assume that $v$ is neither holomorphic nor antiholomorphic. According to (\ref{isotropy}), $(v, D_z v, D_{\bar{z}} v)$ and $(v, D_z^2 v, D_{\bar{z}} v)$ are two triples of mutually orthogonal vectors. Since the ambient space is three-dimensional, we have:
\bea\label{2derprop}
D_z^2 v = \beta D_z v
\eea
for some scalar function $\beta$. Upon the introduction of a unit vector $w=\frac{D_z v}{\|D_z v\|}$, a direct calculation shows that this equality may be rewritten as follows:
\bea\label{antihol}
D_z^{(w)}\,w=0
\eea
which implies that $w=\frac{D_z v}{\|D_z v\|}$ is antiholomorphic. Analogously $\frac{D_{\bar{z}} v}{\|D_{\bar{z}} v\|}$ is holomorphic.

Since (anti)-holomorphic maps are harmonic, both of these maps constitute solutions of (\ref{CP2eom}) as well. In the general case of $\CP^N$ an analogous statement is a consequence of a remarkable fact, namely the existence of a general B\"acklund transformation, producing new solutions of the e.o.m. out of a given one.

\ssubsubsection{The B\"acklund transformation}

A remarkable fact about the equation (\ref{CP2eom}) is that, given a solution $v(z, \bar{z})$, one can generate another solution $w(z, \bar{z})$ via \cite{Wood}
\bea\label{back1}
w=B\circ v=\frac{D_z v}{\|D_z v\|}
\eea
To see this, note the following fact. If $v$ is a solution of (\ref{CP2eom}), then
\bea\label{covrel}
D_{z}^{(v)}
=f^{-1}\cdot D_z^{(w)}\cdot f,\quad\quad D_{\bar{z}}^{(v)}=f\cdot D_{\bar{z}}^{(w)}\cdot f^{-1},\quad\quad f= \|D_z v\|
\eea
Therefore (\ref{CP2eom}) implies $D_{\bar{z}}^{(w)} w=-f \, v$. Acting by $D_z^{(w)}$ and using (\ref{covrel}), one obtains
\bea
D_z^{(w)}D_{\bar{z}}^{(w)} w=-f^2\,w,
\eea
which means that $w$ is harmonic. (Here we use the second form (\ref{harm2}) of the $\sigma$-model e.o.m.)

Analogously to (\ref{back1}), one can construct a second B\"acklund transform:
\bea
\tilde{w}=\tilde{B}\circ v=-\frac{D_{\bar{z}}v}{\|D_{\bar{z}}v\|}
\eea
It is, in fact, inverse to $B$, when acting on non-(anti)-holomorphic maps:
\bear\label{Binv1}
\tilde{B}\circ B=\mathbf{1}\quad &&\textrm{on non-anti-holom.}\;\;(B\circ v\neq0)\\ \label{Binv2}
B\circ \tilde{B}=\mathbf{1}\quad &&\textrm{on non-holom.}\;\;(\tilde{B}\circ v\neq 0)
\eear
To prove (\ref{Binv1})-(\ref{Binv2}), one should use (\ref{covrel}) and the analogous relations
\bea\label{covrel2}
D_{z}^{(v)}
=\tilde{f}\cdot D_z^{(\tilde{w})}\cdot \tilde{f}^{-1},\quad\quad D_{\bar{z}}^{(v)}=\tilde{f}^{-1}\cdot D_{\bar{z}}^{(\tilde{w})}\cdot \tilde{f},\quad\quad \tilde{f}= \|D_{\bar{z}} v\|
\eea

\vspace{0.3cm}
Combining the results of the discussion above, we arrive at the conclusion that harmonic maps $\CP^1 \to \CP^2$ are generically in $3:1$ correspondence with holomorphic maps $\CP^1 \to \CP^2$. Namely, for \emph{every} holomorphic map $v$ we can construct two additional harmonic descendants: $w_1=B\circ v$ and $w_2=B\circ B\circ v$, the second one being anti-holomorphic, so that 
$B\circ w_2=0$. In the special case when $v$ is not a \emph{full} map, i.e. when it is a map to a proper linear subspace $\CC^2\subset \CC^3$, it turns out that $w_2\equiv 0$, so that there is a single descendant $w_1$, which in this case is anti-holomorphic. The extreme case $w_1\equiv 0$ corresponds to a constant map $v$.

\ssubsection{Lift to the flag manifold}

In order to convert a harmonic map 
\bea\label{harmu2}
v=u_2: \CP^1 \to \CP^2
\eea
 into a critical map to $\mathcal{F}_3$, we wish to show that we can lift the former to the flag manifold, satisfying the remaining equation (\ref{J31}): $(J_{31})_{\bar{z}}=u_3 \circ \dd_{\bar{z}}\bar{u}_1=0$, where $u_1$ and $u_3$ are orthogonal to each other and to $u_2$.

\vspace{0.7cm}
\textbf{I. $D_z u_2\nequiv0, D_{\bar{z}}u_2 \nequiv 0$}. Both of these vectors are orthogonal to $u_2$ (by definition) and to each other (by the isotropy property). Therefore $u_1$ and $u_3$ are linear combinations of these two vectors:
\bear
&&u_1= a \,D_zu_2 +b \,D_{\bar{z}}u_2,\\
&&u_3= c \,D_zu_2+d \,D_{\bar{z}}u_2
\eear
Acting on $u_3$ with $D_{\bar{z}}^{(u_2)}$, we obtain ($\alpha$ is the scalar function from (\ref{u2harm})):
\bea
D_{\bar{z}}^{(u_2)} u_3= \dd_{\bar{z}}c\,D_zu_2+c \,\alpha\, u_2+(\dd_{\bar{z}}\, d+\tau \,d)\,D_{\bar{z}}u_2,
\eea
where $\tau$ is the proportionality constant from the equality $(D_{\bar{z}}^{(u_2)})^2 u_2=\tau D_{\bar{z}} u_2$ (which is derived analogously to (\ref{2derprop})). A simple calculation shows that
\bea
\tau=\dd_{\bar{z}}(\log\|D_{\bar{z}}u_2\|^2)\, .
\eea
The equation $u_3 \circ \dd_{\bar{z}}\bar{u}_1=0$ then requires
\bea
\dd_{\bar{z}}c\,\bar{a}\,\|D_zu_2\|^2+\bar{b}\,(\dd_{\bar{z}}\, d+\tau\, d)\,\|D_{\bar{z}}u_2\|^2=0
\eea
Together with the orthogonality condition $\bar{u}_1\circ u_3=0$, expressed as
\bea\label{abcd}
c\,\bar{a}\,\|D_zu_2\|^2+\bar{b}\, d\,\|D_{\bar{z}}u_2\|^2=0,
\eea
this leads to
\bea\label{cdeq}
\dd_{\bar{z}}c\, d- (\dd_{\bar{z}} d+\tau\, d)\, c=0,
\eea
hence
\bea
\left(\begin{array}{c}c  \\d \end{array}\right)=\lambda(z, \bar{z})\,\left(\begin{array}{c}f(z)\cdot\|D_{\bar{z}}u_2\|^2  \\g(z) \end{array}\right),
\eea
with two holomorphic functions $(f(z) : g(z)) \in \CP^1$. It is easy to see that the remaining unknowns, such as $\lambda, a, b$ can be now found from the normalization conditions $\bar{u}_1 \circ u_1=\bar{u_3}\circ u_3=1$. Therefore what defines the lift to the flag manifold is a holomorphic map
\bea\label{fg}
(\CP^1)_z \to (\CP^1)_{(f:g)}\;.
\eea
One can also think of this map as a rational function $\frac{f(z)}{g(z)}$.

Note that the critical map $\CP^1\to \mathcal{F}_3$ constructed in this fashion is not holomorphic in either of the almost complex structures on $\mathcal{F}_3$, unless the matrix $\left(\begin{array}{cc}a & b  \\c & d\end{array}\right)$ has some zero elements. This is so, since $D_z u_2, D_{\bar{z}}u_2$ are not orthogonal to either $u_1$ or $u_3$, hence violating the holomorphicity conditions for all complex structures. Due to (\ref{abcd}), the only possibilities for the above matrix to have zero elements are as follows:

\vspace{0.3cm}
\textbf{Ia.} $a=d=0$, i.e. $u_1\sim D_{\bar{z}}u_2,\; u_3\sim D_z u_2$. Then $\bar{u}_1\circ D_z u_2=0=\bar{u}_3\circ D_{\bar{z}}u_2$. It is also easy to check that $\bar{u}_3\circ D_z u_1=0$, as well as $\bar{u}_3 \circ D_{\bar{z}}u_1=0$. This means that the lift is a horizontal curve (with respect to the twistor fibration), which is holomorphic in complex structures $\mathcal{Q}_1$ and $\mathcal{Q}_I$.

\textbf{Ib.} $b=c=0$, i.e. $u_3\sim D_{\bar{z}}u_2,\; u_1\sim D_z u_2$. This is essentially a $u_1 \leftrightarrow u_3$ reversal of the case \textbf{Ia}. Therefore the lift in this case is a horizontal curve, holomorphic in $\mathcal{Q}_{-1}$ and $\mathcal{Q}_{-I}$. Note that this is an exceptional case when the curve is holomorphic in the complex structure $\mathcal{Q}_{-1}$, not shown in Fig. \ref{cstr}. Such holomorphicity is possible due to the horizontality of the map, i.e. $J_{13}\equiv 0$.

\vspace{0.7cm}
\textbf{II. $D_{\bar{z}}u_2=0$.} In this case $u_2$ is a holomorphic map. The condition is equivalent to the following two:
\bea\label{hol1}
(J_{21})_{\bar{z}}=0,\quad\quad (J_{23})_{\bar{z}}=0
\eea
The remaining e.o.m, (\ref{J31}), states that
\bea\label{hol2}
(J_{31})_{\bar{z}}=0\,.
\eea
Together the above equations (\ref{hol1})-(\ref{hol2}) imply that we are dealing with a curve $\mathscr{M}~\to~\mathcal{F}_3$, holomorphic in the complex structure, defined by the graph $\mathcal{Q}_3$.

\vspace{0.7cm} \textbf{III. $D_{z}u_2=0$.} Hence $u_2$ is an anti-holomorphic map. In this case
\bea
(J_{12})_{\bar{z}}=0,\quad\quad (J_{32})_{\bar{z}}=0,\quad\quad (J_{31})_{\bar{z}}=0
\eea
which corresponds to a curve, holomorphic in the complex structure $\mathcal{Q}_2$. 

\vspace{0.3cm} The situation when $D_{\bar{z}}u_2=D_{z}u_2=0$, i.e. when $u_2$ is a map to a point, is at the intersection of cases \textbf{II} and \textbf{III}. In this case, due to the condition $(J_{31})_{\bar{z}}=0$, $(u_1, u_3)$ specify a holomorphic map to a $\CP^1$, orthogonal to the fixed vector $u_2$. In other words, it is a map to the fiber of the fibration $\mathcal{F}_3\to (\CP^2)_{u_2}$, and this map is holomorphic in two complex structures, $\mathcal{Q}_2$ and $\mathcal{Q}_3$. This property was already observed in \cite{Bykov}.

Analysis of the cases, when in place of (\ref{J31}) one has $(J_{12})_{\bar{z}}=0$ or $(J_{23})_{\bar{z}}=0$, goes along the same lines, with obvious permutations of $u_1, u_2, u_3$.

\ssection{Summary}

In this paper we have solved the e.o.m. (\ref{eomflag}), which follow from the action (\ref{action}), introduced in \cite{Bykov}. The solutions that we obtained correspond to the case when the worldsheet is the sphere $\CP^1$, and they exhaust all solutions in this case. We have shown, that, apart from various holomorphic curves, there exists a subclass of solutions that are not holomorphic in any (almost) complex structure on $\mathcal{F}_3$. The data for such solutions consist of a full holomorphic curve $\CP^1\to\CP^2$ -- the `B\"acklund primitive' of (\ref{harmu2}) -- and a holomorphic map $\CP^1\to\CP^1$ (\ref{fg}).

The key property which allowed us to solve the equations is that, due to the fact that $\CP^1$ does not have holomorphic differentials, the problem reduced to the one of finding harmonic curves in $\CP^2$ (see (\ref{noholdif})-(\ref{u2harm})), and the latter problem was solved long ago \cite{Din}. This approach is not directly generalizable to other worldsheets. However, in \cite{Bykov} it was shown that the e.o.m. (\ref{eomflag}) can be written in terms of a one-parametric family of flat connections. For $\sigma$-models with symmetric target spaces such representation provides a method for the construction of solutions, which  was developed in \cite{Zakharov} and rigorously justified in \cite{Uhlenbeck}. It would be very interesting to explore, whether a suitable modification of the method would allow to obtain all solutions of the equations (\ref{eomflag}) in the case when the worldsheet is not a sphere but rather a higher-genus Riemann surface, or a cylinder. 

\vspace{0.3cm}
\textbf{Acknowledgements.}
{\footnotesize
I am indebted to Prof.~A.A.Slavnov and to my parents for support and encouragement. I would like to thank the Institut des Hautes \'Etudes Scientifique (IHES), where part of the work was done, and especially V.~Pestun for hospitality. My work was supported in part by grants RFBR 14-01-00695-a, 13-01-12405 ofi-m2 and the grant MK-2510.2014.1 of the President of Russia Grant Council.
}

\vspace{-1.3cm}
\renewcommand\refname{\begin{center} \centering\normalfont\scshape  References\end{center}}


\end{document}